\documentclass{SciPost}

\hypersetup{
    colorlinks,
    linkcolor={red!50!black},
    citecolor={blue!50!black},
    urlcolor={blue!80!black}
}

\usepackage[bitstream-charter]{mathdesign}
\usepackage{booktabs}
\usepackage[export]{adjustbox}
\usepackage{subcaption}

\usepackage{tabularx}

\DeclareSymbolFont{usualmathcal}{OMS}{cmsy}{m}{n}
\DeclareSymbolFontAlphabet{\mathcal}{usualmathcal}

\fancypagestyle{SPstyle}{
\fancyhf{}
\lhead{\colorbox{scipostblue}{\bf \color{white} ~SciPost Physics Community Reports }}
\rhead{{\bf \color{scipostdeepblue} ~Submission }}

\fancyfoot[C]{\textbf{\thepage}}
}

\newcommand{\mhh}{m_{hh}}

\usepackage{comment}

\usepackage[dvipsnames]{xcolor}

\begin{document}

\pagestyle{SPstyle}
%
% preprint numbers
%\begin{flushright}\footnotesize{KA-TP-xx-2026, P3H-26-yy}\end{flushright}
%
%
\begin{center}{\Large \textbf{\color{scipostdeepblue}{
%%%%%%%%%% TODO: Write your article's title here
        Higgs boson pair production in SMEFT: shape and truncation studies
}}}\end{center}

\begin{center}\textbf{
% Use (full) first name (+ middle name initials) + surname format.
% Separate subsequent authors by a comma, omit comma and use "and" for the last author.
% Mark the corresponding author(s) with a superscript symbol in this order
% \star, \dagger, \ddagger, \circ, \S, \P, \parallel, ...
Manosch B\"ar\textsuperscript{1,2},
Bakar Chargeishvili\textsuperscript{1},
Ramona Gr\"ober\textsuperscript{3},
Gudrun Heinrich\textsuperscript{1},
Konstantin Schmid\textsuperscript{3}
}\end{center}

\begin{center}
%%%%%%%% %% TODO: AFFILIATIONS
{\bf 1} Institute for Theoretical Physics, Karlsruhe Institute
  of Technology (KIT), 76131 Karlsruhe, Germany
\\
{\bf 2}  Institute for Mathematics, Astrophysics and Particle Physics, Radboud University Nijmegen, 6525 AJ Nijmegen, The Netherlands \\
{\bf 3}  Dipartimento di Fisica e Astronomia ``G. Galilei'', Universit\`a di Padova and Istituto Nazionale di Fisica Nucleare, Sezione di Padova, I-35131 Padova, Italy
% 
%%%%%%%%%% END TODO: AFFILIATIONS
% Provide email address of corresponding author(s)
\\[\baselineskip]
\href{mailto:manosch.baer@kit.edu}{\small manosch.baer@partner.kit.edu},
\href{mailto:bakar.chargeishvili@kit.edu}{\small bakar.chargeishvili@kit.edu},  
\href{mailto:ramona.groeber@pd.infn.it}{\small 
ramona.groeber@pd.infn.it},
\href{mailto:gudrun.heinrich@kit.edu}{\small gudrun.heinrich@kit.edu}, 
\href{mailto:konstantin.schmid@pd.infn.it}{\small konstantin.schmid@pd.infn.it}
%$\dagger$ \href{mailto:email2}{\small email2}
\end{center}

\section*{\color{scipostdeepblue}{Abstract}}
\textbf{\boldmath{We present a comparative study of the influence of Wilson coefficients on Higgs boson pair production cross sections in SMEFT up to dimension-6, also including loop-suppressed operators. Our study accounts for NLO QCD corrections, and is based on updated parametrizations of the cross section as a polynomial in the Wilson coefficients. We focus on differential distributions for the Higgs boson pair invariant mass, as well as the transverse momenta of the leading Higgs bosons, identifying benchmark scenarios exhibiting pronounced shape distortions in the $m_{hh}$ and $p_{T,h}$ distributions. While some of these scenarios remain physically meaningful under a purely linear truncation of the SMEFT expansion, others yield unphysical differential cross sections when quadratic dimension-6 contributions are omitted, exposing the breakdown of the linear approximation for large anomalous couplings.}}

\vspace{\baselineskip}

%%%%%%%%%% BLOCK: Copyright information
% This block will be filled during the proof stage, and finilized just before publication.
% It exists here only as a placeholder, and should not be modified by authors.
\noindent\textcolor{white!90!black}{%
\fbox{\parbox{0.975\linewidth}{%
\textcolor{white!40!black}{\begin{tabular}{lr}%
  \begin{minipage}{0.6\textwidth}%
    {\small Copyright attribution to authors. \newline
    This work is a submission to SciPost Phys. Comm. Rep. \newline
    License information to appear upon publication. \newline
    Publication information to appear upon publication.}
  \end{minipage} & \begin{minipage}{0.4\textwidth}
    {\small Received Date \newline Accepted Date \newline Published Date}%
  \end{minipage}
\end{tabular}}
}}
}\newpage
%%%%%%%%%% BLOCK: Copyright information

% For convenience during refereeing we turn on line numbers:
%\linenumbers

\vspace{10pt}
\noindent\rule{\textwidth}{1pt}
\tableofcontents
\noindent\rule{\textwidth}{1pt}
\vspace{10pt}

%\newpage
% Write your article contents here, starting from first \section.
% An example structure is given below.

\section{Introduction}
Di-Higgs production from gluon fusion offers the unique opportunity to directly probe the trilinear Higgs self-interaction, thereby accessing the structure of the Higgs potential, as well as possible anomalous couplings of the Higgs boson to gluons and top quarks \cite{DiMicco:2019ngk}. Deviations from the Standard Model (SM) prediction induced by heavy new physics (NP) can be systematically captured within an effective field theory (EFT) framework and manifest themselves in modifications of both the total cross section and the shapes of differential distributions. While such shape changes have been extensively studied in the Higgs Effective Field Theory (HEFT) framework \cite{Carvalho:2015ttv, Buchalla:2018yce, Capozi:2019xsi, Brivio:2025sib}, they remain comparatively unexplored in the context of the Standard Model Effective Field Theory (SMEFT).

This article aims to provide a systematic phenomenological study in the essential kinematic observables of di-Higgs production within the SMEFT up to dimension-6, including next-to-leading order (NLO) QCD corrections. In particular, we study the invariant mass distributions of the Higgs boson pair as well as the leading transverse momentum spectra. We further examine the impact of the SMEFT truncation scheme, comparing linear and quadratic treatments, and assess the associated truncation uncertainties in light of the validity of the SMEFT expansion.

This paper is structured as follows. In Sec.~\ref{sec_SMEFT}, we briefly review the SMEFT framework and establish our notation and conventions for the dimension-6 operators. Furthermore, we discuss the procedure and give the input values of our numerical study. Sec.~\ref{sec_pheno} presents our phenomenological results for the differential di-Higgs shapes and introduces benchmark points for the invariant mass. This is followed by a discussion of the impact of the adopted truncation procedure. Finally, we summarize our findings and conclude in Sec.~\ref{sec_conclusions}.

\section{SMEFT} \label{sec_SMEFT}
This section is dedicated to a brief summary of our conventions and the setup we use for our analysis of di-Higgs production in the SMEFT. For comprehensive reviews on SMEFT we refer to e.g. Refs.~\cite{Brivio:2017vri, Isidori:2023pyp, Aebischer:2025qhh}. The SMEFT extends the SM by introducing operators with mass dimension greater than four that respect the symmetries of the SM, adopting the $SU(2)$ doublet representation of the Higgs field $H$. Its expansion is organized in terms of the canonical dimension $d$ of the effective operators as
\begin{align}
    \mathcal{L}_{\rm SMEFT} = \mathcal{L}_{\rm SM} + \sum_{d>4} \sum_i C_i^{(d)} \frac{\mathcal{O}_i^{(d)}}{\Lambda^{d-4}} \, ,
\end{align}
where $\mathcal{L}_{\rm SM}$ is the renormalizable SM Lagrangian, the $C_i^{(d)}$ are dimensionless Wilson coefficients, and $\Lambda$ is a heavy mass scale that ensures the correct mass dimension for all terms in $\mathcal{L}_{\rm SMEFT}$. Given that the SMEFT expansion is controlled by inverse powers of $\Lambda$, a SMEFT prediction for a cross section can be written as
\begin{align}
    \label{SMEFT_expansion}
    \sigma_{\rm SMEFT} = \sigma_{\rm SM} + \sum_i \frac{C_i^{(6)}}{\Lambda^2} \sigma^{(\rm SM \times 6)}_i + \sum_{\substack{i,j \\ i \le j}} \frac{C_i^{(6)} C_j^{(6)}}{\Lambda^4} \sigma^{(6 \times 6)}_{ij} + \sum_i \frac{C_i^{(8)}}{\Lambda^4} \sigma^{(\rm SM \times 8)}_i + \mathcal{O}(\Lambda^{{-6}}) \, ,
\end{align}
where we spelled out all contributions up to $\mathcal{O}(\Lambda^{{-4}})$. In the context of this article, we will focus on the $d=6$ operators relevant for the production of a Higgs boson pair from gluon fusion that can be parametrized using the Warsaw basis \cite{Grzadkowski:2010es} as
\begin{equation}
\begin{aligned}
    \mathcal{L}_{\rm SMEFT}^{(6)} \supset &\frac{C_{H \Box}}{\Lambda^2} (H^{\dagger} H) \Box (H^{\dagger} H) + \frac{C_{HD}}{\Lambda^2} (H^{\dagger} D_{\mu} H)^{\ast} (H^{\dagger} D^{\mu} H) + \frac{C_H}{\Lambda^2} (H^{\dagger} H)^3 \\
    &+ \frac{C_{tH}}{\Lambda^2} H^{\dagger} H (\bar{Q}_L \tilde{H} t_R + \mathrm{h.c.}) + \frac{C_{HG}}{\Lambda^2} H^{\dagger} H G_{\mu \nu}^a G^{a\mu \nu} \\
    &+ \frac{C_{tG}}{\Lambda^2} (\bar{Q}_L \sigma^{\mu \nu} T^a G^a_{\mu \nu} \tilde{H} t_R + \mathrm{h.c.}) \\
    &+ \frac{C_{Qt}^{(1)}}{\Lambda^2} \bar{Q}_L \gamma^{\mu} Q_L \bar{t}_R \gamma_{\mu} t_R + \frac{C_{Qt}^{(8)}}{\Lambda^2} \bar{Q}_L \gamma^{\mu} T^a Q_L \bar{t}_R \gamma_{\mu} T^a t_R \, .
\label{Lagrangian_SMEFT}
\end{aligned}
\end{equation}
Setting aside scenarios with significantly enhanced light-quark Yukawa couplings~\cite{Alasfar:2019pmn, Egana-Ugrinovic:2021uew, Alasfar:2022vqw}, we only include the top-quark, as the contribution of the light quarks is expected to be Yukawa-suppressed to this order. However, we do not discuss the same chirality four-fermion operators $Q_{QQ}^{(1)}, Q_{QQ}^{(8)}$, and $Q_{tt}$, since their numerical contribution to Higgs pair production has been shown to be suppressed \cite{Heinrich:2023rsd}. After spontaneous breaking of the $SU(2)_L \times U(1)_Y$ symmetry, $Q_{H \Box}$ and $Q_{HD}$ give rise to additional pieces in the kinetic term of the physical Higgs field $h$. A field redefinition of the form
\begin{align}
    h \to h + v^2 \frac{C_{H, \rm kin}}{\Lambda^2} \left(h + \frac{h^2}{v}+\frac{h^3}{3 v^2} \right)
\end{align}
with $C_{H, \mathrm{kin}} = C_{H\Box}-C_{HD}/4$ restores canonical normalization and removes the momentum-dependent structures in the Higgs self-interactions. 

\subsection{SMEFT Truncation}
In practice, the SMEFT series expansion as written in Eq.~(\ref{SMEFT_expansion}) has to be truncated to perform theoretical predictions up to a fixed order. This, however, leads to the question of whether dimension-6 contributions should be retained only at linear order, or whether quadratic terms in the dimension-6 coefficients should also be included, despite being formally of order $\mathcal{O}(\Lambda^{-4})$ at the level of the squared amplitude. This issue, along with the estimation of the corresponding truncation error, has been subject to discussions in the recent literature, see, for instance, Refs.~\cite{Brivio:2022pyi, Trott:2021vqa, Chang:2025ohh}. In this work, we will compare the previously mentioned truncation approaches. First, we study a linear truncation that strictly disregards all terms of order $\mathcal{O}(\Lambda^{{-4}})$, leading to a parametrization 
\begin{align}
    \sigma_{\rm SMEFT}\big|_{\rm lin} = \sigma_{\rm SM} + \sum_i A_i^{\rm incl} \, C_i^{(6)}
\label{param_lin_incl}
\end{align}
of the total SMEFT cross section and 
\begin{align}
    \frac{d \sigma_{\rm SMEFT}}{d O}\bigg|_{\rm lin} = \frac{d \sigma_{\rm SM}}{d O} + \sum_i A_i[O] \, C_i^{(6)}
\label{param_lin_diff}
\end{align}
for differential distributions in the observable $O$. As a second option, we also include the terms proportional to the product of two dimension-6 coefficients, which stem from the square of the dimension-6 amplitude. We do not include dimension-6 double insertions as they do not form a closed set under field redefinitions and renormalization that is fully disconnected from dimension-8 operators \cite{Helset:2022pde, Chala:2021pll, Chala:2021wpj, DasBakshi:2022mwk, Wu:2025qto, Heinrich:2022idm, Martin:2021cvs}.
Nevertheless, we note that the squared dimension-6 contributions formally enter at the same EFT order as the interference of dimension-8 operators with the SM amplitude. Terming this approach quadratic truncation, we write
\begin{align}
    \sigma_{\rm SMEFT}\big|_{\rm quad} &= \sigma_{\rm SMEFT}\big|_{\rm lin} + \sum_{\substack{i,j \\ i \le j}} A_{ij}^{\rm incl} \, C_i^{(6)} C_j^{(6)} \, ,
\label{param_quad_incl}
\end{align}
and
\begin{align}
    \frac{d \sigma_{\rm SMEFT}}{d O}\bigg|_{\rm quad} &= \frac{d \sigma_{\rm SMEFT}}{d O}\bigg|_{\rm lin} + \sum_{\substack{i,j \\ i \le j}} A_{ij}[O] \, C_i^{(6)} C_j^{(6)} \, .
\label{param_quad_diff}
\end{align}
Clearly, all predictions in the quadratic framework originate from
\begin{align}
     |\mathcal{M}|^2 = \left|\mathcal{M}_{\rm SM} + \mathcal{M}^{(6)}\right|^2
\end{align}
and are therefore strictly positive, whereas linear truncation may lead to a negative inclusive cross section or negative bins in the differential distributions, possibly signalizing a region in the parameter space in which the EFT is not valid.

\subsection{Determination of $A_i$-coefficients and Inputs}
The $A_i$-coefficients in SMEFT including NLO QCD contributions were obtained using an iterative procedure. Each iteration starts with Latin Hypercube Sampling (LHS)~\cite{mckay:1979}, which divides the Wilson coefficient parameter space into $N$ regions, where $N$ is the number of samples. Each region is sampled randomly, ensuring uniform projection onto each parameter while maximizing spread and avoiding clustering. The \texttt{ggHH\_SMEFT} package \cite{Heinrich:2022idm, Heinrich:2023rsd} in \texttt{POWHEG-BOX-V2} \cite{Nason:2004rx, Frixione:2007vw, Alioli:2010xd}, with the inputs from Tab.~\ref{Tab_inputs}, is used to compute the data.

Each variable - the seven Wilson coefficients and the cross section - of each data point is assigned a label based on the quantile its value falls into. The number of quantiles is determined by $N_{\rm data}/N_{\rm folds}$ rounded down, where $N_{\rm folds} \in \mathbb{N}$ is the number of folds, defined as the ratio $N_{\rm data}/N_{\rm validation \ data}$. A multilabel stratification algorithm is then used to split the data into two subsets where one is used for fitting and another is used for validation, ensuring that the individual distributions of each variable are preserved between both sets.

Fitting is performed via least squares, with uncertainties for the $A_i$-coefficients derived from the diagonal of the covariance matrix. The reduced-$\chi^2$ metric is used to compare the score from the set used for fitting against a validation set. Iterations continue until the reduced-$\chi^2$ score of the sets used for fitting and validation converge to the same score, indicating that fit has generalized and more iterations are unlikely to improve the accuracy. The results for the differential and inclusive $A_i$-coefficients will be provided in 
a separate contribution to Report~5~\cite{Baer:2026}, 
following the parameterization of App.~\ref{appendix_parameterization}.

\begin{table}[t]
\centering
\begin{tabular}{c|c|c}
\hline
Input & Value & Description \\ \hline
$m_t$ & $173.0 \, \mathrm{GeV}$ & mass of the top-quark \\
$m_Z$ & $91.1876 \, \mathrm{GeV}$ & mass of the Z-boson \\
$m_h$ & $125.0 \, \mathrm{GeV}$ & mass of the Higgs boson \\
$G_F$ & $1.16639 \times 10^{-5} \, \mathrm{GeV}^{-2}$ & Fermi's constant \\
PDF & \texttt{PDF4LHC21\_40} & parton distribution functions \\
$\alpha_s(m_Z)$ & 0.118 & strong coupling at the $Z$-pole \\
$\sqrt{s}$ & $13.6 \, \mathrm{TeV}$ & COM energy \\
$\Lambda$ & $1 \, \mathrm{TeV}$ & SMEFT NP scale \\
$\bar{\mu}$ & $m_{hh}/2$ & central renormalization and factorization scale \\ 
\hline
\end{tabular}
\caption{Inputs for our numerical study of $pp \to hh$.}
\label{Tab_inputs}
\end{table} 

\section{Results and Phenomenology} \label{sec_pheno}
In our phenomenological analysis of the di-Higgs invariant mass and transverse momentum distributions, we apply bounds on the dimension-6 Wilson coefficients that stem from single-Higgs production and top quark physics, including the mixing with other operators via the renormalization group equations (RGEs). The results of recent global analyses and phenomenological studies are summarized in Tab.~\ref{Tab_SMEFT_bounds}.

\begin{table}[t]
\centering
\renewcommand{\arraystretch}{1.2}
\begin{tabular}{c|cc|cc}
\hline
\text{Wilson coefficient} & \multicolumn{2}{c|}{\text{bound linear}} & \multicolumn{2}{c}{\text{bound quadratic}} \\
\hline
$C_{H\Box}$ & $[-2.95, 2.20]$ & \cite{terHoeve:2025gey} & $[-2.15, 1.50]$ & \cite{terHoeve:2025gey} \\
$C_{HD}$ & $[-3.01, 6.05]$ & \cite{terHoeve:2025gey} & $[-0.24, 1.15]$ & \cite{terHoeve:2025gey} \\
$C_{H}$ & $[-9.21, 8.61]$ & \cite{terHoeve:2025gey} & $[-12.52, 3.78]$ & \cite{terHoeve:2025gey} \\
$C_{tH}$ & $[-15.18, 3.45]$ & \cite{terHoeve:2025gey} & $[-4.15, 4.38]$ & \cite{terHoeve:2025gey} \\
$C_{HG}$ & $[-0.04, 0.02]$ & \cite{terHoeve:2025gey} & $[-0.03, 0.00]$ & \cite{terHoeve:2025gey} \\
$C_{tG}$ & $[-0.02, 0.27]$ & \cite{terHoeve:2025gey} & $[0.00, 0.18]$ & \cite{terHoeve:2025gey} \\
$C_{Qt}^{(1)}$ & $[-10.99, 6.71]$ & \cite{DiNoi:2025uhu} & $[-1.37, 2.69]$ & \cite{DiNoi:2025uhu} \\
$C_{Qt}^{(8)}$ & $[-9.50, 58.36]$ & \cite{DiNoi:2025uhu} & $[-4.09, 6.20]$ & \cite{DiNoi:2025uhu} \\
\hline
\end{tabular}
\caption{Bounds on dimension-6 SMEFT Wilson coefficients that enter our study of di-Higgs production. The constraints from Ref.~\cite{terHoeve:2025gey} are obtained using marginalization from a global fit, while Ref.~\cite{DiNoi:2025uhu} marginalizes over the four-top quark operators.}
\label{Tab_SMEFT_bounds}
\end{table}
Given the constraints on $C_{H\Box}$ and $C_{HD}$, we can translate them into a bound on the linear combination $C_{H, \mathrm{kin}}$ that enters our discussions. Assuming independent variation of $C_{H \Box}$ and $C_{HD}$, and neglecting their correlations, we obtain
\begin{align}
    C_{H, \mathrm{kin}} \in [-4.46, 2.95]_{\mathrm{lin}} \ \, \mathrm{and} \ \, [-2.44, 1.56]_{\mathrm{quad}} \, .
\end{align}
In addition, we require that the total SMEFT cross section should not exceed a rather conservative bound on the signal strength; we choose this bound as $\mu_{hh} < 8$, which is motivated by the experimental analyses in the leading $\bar{b}b\bar{b}b, \bar{b}b \tau^{+} \tau^{-}$, and $\bar{b}b \gamma \gamma$ Higgs decay channels by the ATLAS and CMS collaborations~\cite{CMS:2026nuu, ATLAS:2023qzf, ATLAS:2024lsk, ATLAS:2024pov, ATLAS:2023gzn, CMS:2022cpr, CMS:2022gjd, CMS:2022hgz, CMS:2020tkr}. 

\subsection{Invariant Mass Benchmarks}
\begin{figure}[t]
    \centering
    \includegraphics[width=0.7\linewidth]{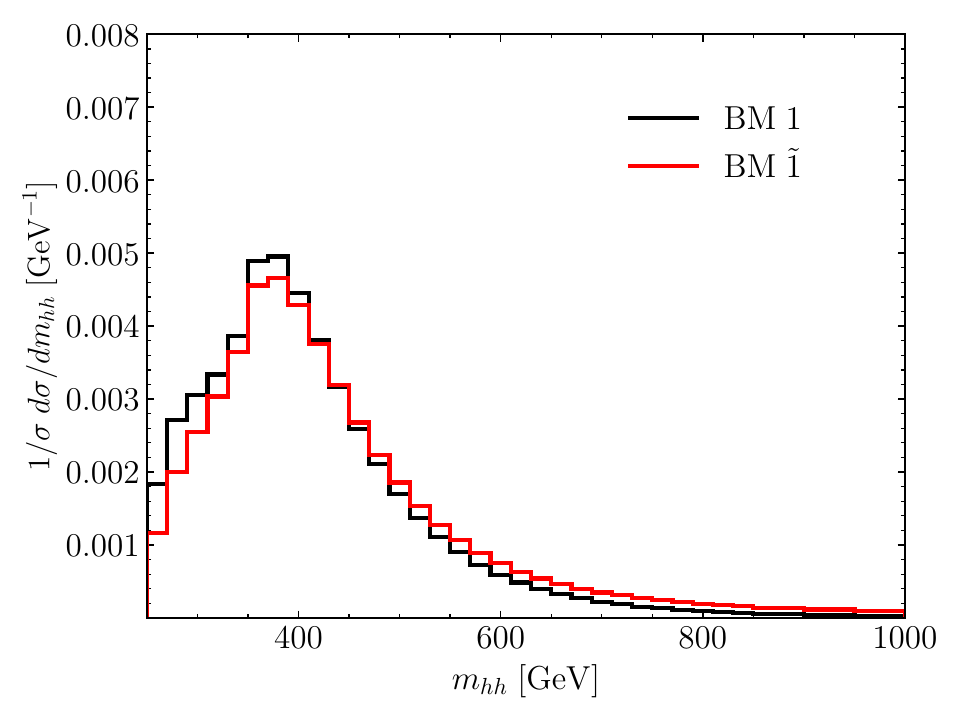}
    \caption{Benchmark points $1$ and $\tilde{1}$ for quadratic truncation. Both distributions exhibit a shoulder-like kinematic feature, exemplifying the degeneracy.}
    \label{fig_degeneracy}
\end{figure}

\begin{figure}[t]
    \centering
    \includegraphics[width=1.0\linewidth]{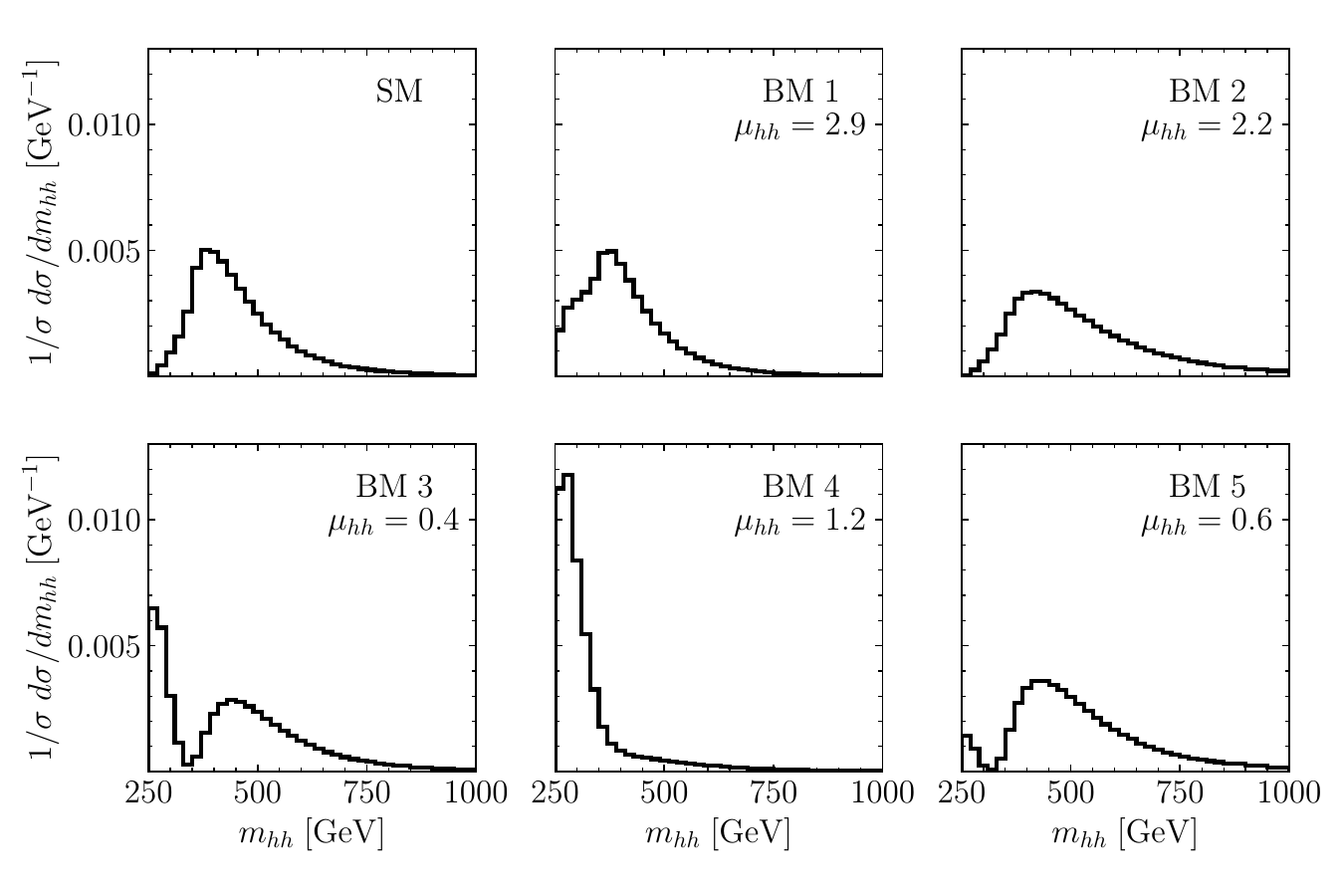}
    \caption{Benchmark points for dimension-6 SMEFT with quadratic truncation.}
    \label{benchmarks_quadratic}
\end{figure}

\noindent In Tab.~\ref{benchmarks_invmass} we present benchmark points for the quadratic truncation, which lead to distinct kinematic features in the $m_{hh}$-shapes. 
\begin{table}[h]
\centering
\begin{tabular}{c|c|c|c|c|c|c|c|c}
\hline
Benchmark (BM) & $C_{H, \mathrm{kin}}$ & $C_{H}$ & $C_{tH}$ & $C_{HG}$ & $C_{tG}$ & $C_{Qt}^{(1)}$ & $C_{Qt}^{(8)}$ & $\mu_{hh}$ \\ \hline
SM & 0 & 0 & 0 & 0 & 0 & 0 & 0 & 1 \\
1 & 0 & $3$ & 0 & 0 & 0 & 0 & 0 & 2.9  \\
2 & $-2$ & $-2$ & $4$ & 0 & 0 & 0 & 0 & 2.1 \\
3 & 0 & $-3$ & 0 & 0 & 0 & 0 & 0 & 0.4 \\
4 & 0 & $-6$ & 0 & 0 & 0 & 0 & 0 & 1.2 \\
5 & $-2$ & $-3$ & $1/2$ & 0 & 0 & 0 & 0 & 0.6 \\
\hline
\end{tabular}
\caption{Benchmark (BM) points for dimension-6 SMEFT with quadratic truncation and $\Lambda = 1\, \mathrm{TeV}$. In the limit $\Lambda\to \infty$ all shapes converge to the SM.}
\label{benchmarks_invmass}
\end{table} \\
While in general similar kinematic features can be obtained for different choices of Wilson coefficients, we determined our selection of benchmark points such that a minimal set of non-zero coefficients is involved. The main source of deviations from the SM is the coefficient $C_H$, as its only direct probe is Higgs pair production, and thus it remains weakly constrained. Furthermore, it influences delicate interference patterns in the amplitude. In contrast, the full shape information can be obtained by setting the coefficients $C_{HG}, C_{tG}, C_{Qt}^{(1)}$, and $C_{Qt}^{(8)}$ to zero. The primary effect of $C_{HG}$ are shoulder-like features, which are degenerate with respect to \mbox{BM 1}. In particular, one could define a BM $\tilde{1}$ as 
\begin{table}[h]
\centering
\begin{tabular}{c|c|c|c|c|c|c|c|c}
Benchmark (BM) & $C_{H, \mathrm{kin}}$ & $C_{H}$ & $C_{tH}$ & $C_{HG}$ & $C_{tG}$ & $C_{Qt}^{(1)}$ & $C_{Qt}^{(8)}$ & $\mu_{hh}$ \\ \hline
$\tilde{1}$ & 0 & 0 & 0 & $-0.03$ & 0 & 0 & 0 & $2.8$
\end{tabular}
\end{table} \\
leading to the normalized distribution shown in Fig.~\ref{fig_degeneracy}.
On the other hand, $C_{tG}$ is constrained from single-Higgs and top-quark production and since it enters the di-Higgs amplitude with an additional loop-suppression factor compared to $C_{HG}$, its effect remains relatively small. Furthermore, the coefficients $C_{Qt}^{(1)}$ and $C_{Qt}^{(8)}$ contribute at two-loop level, resulting in a suppression factor of $(16 \pi^2)^{-2}$. In particular, we have explicitly verified that their effects on the normalized distributions remain small when treated in the naive dimensional regularization (NDR) scheme \cite{Chanowitz:1979zu} for the continuation of $\gamma_5$ to $d$ dimensions. In addition, this conclusion also extends to the Breitenlohner--Maison--'t Hooft--Veltman (BMHV) scheme \cite{tHooft:1972tcz,Breitenlohner:1977hr}, since the bounds reported in Tab.~\ref{Tab_SMEFT_bounds} are predominantly driven by four-top production \cite{DiNoi:2025uhu}, for which scheme-dependent differences are irrelevant.
Detailed discussion on the differences between NDR and BMHV scheme in single and double Higgs production can be found in Refs.~\cite{DiNoi:2023ygk, Heinrich:2023rsd}.
The normalized distributions corresponding to the parameters of Tab.~\ref{benchmarks_invmass} are shown in Fig.~\ref{benchmarks_quadratic} and the main kinematic features of the selected benchmarks are summarized in Tab.~\ref{Tab_SMEFT_benchmarks_mhh_quad_features}.
In particular, we observe that these kinematic features resemble the ones that have been identified in the HEFT analysis of the same process \cite{Carvalho:2015ttv, Buchalla:2018yce, Capozi:2019xsi, Brivio:2025sib}.
\begin{table}[t]
\centering
\begin{tabular}{c|c}
\hline
Benchmark (BM) & Kinematic Feature \\ \hline
SM & approximate cancellation at $2m_h$, peak at around $2m_t$ \\
1 & left-sided shoulder with peak at around $2m_t$ \\
2 & smaller peak at $2m_t$ and enhanced tail relative to SM \\
3 & double-peak with dominant (sub-dominant) peak at $2m_h$ ($2m_t$)\\
4 & peak at $2m_h$, essentially flat after $2m_t$ \\
5 & double-peak with dominant (sub-dominant) peak at $2m_t$ ($2m_h$)\\
\hline
\end{tabular}
\caption{Kinematic features of the benchmark points for dimension-6 SMEFT with quadratic truncation.}
\label{Tab_SMEFT_benchmarks_mhh_quad_features}
\end{table}
Finally, we emphasize that our benchmark points do not include operators inducing a power-like growth of the amplitude with energy, such as $O_{HG}$. The operator $O_{tH}$ leads to a logarithmic energy dependence in the high energy limit; however, for the Wilson-coefficient values listed in Table~\ref{benchmarks_invmass}, this growth remains mild and the coefficients lie well below the unitarity limits derived in Ref.~\cite{Bresciani:2026acy}. Moreover, the high-energy region is strongly suppressed by the gluon luminosity. Consequently, the $\mhh$ spectrum does not grow for $\mhh\to 1$\,TeV, such that unitarity violation is not an issue.

\subsection{Renormalization Group Effects}
Since the benchmarks provided in the previous section have been obtained neglecting their running and mixing under the RGEs, we briefly comment on their effect. Clearly, sizeable RGE effects might have an impact on the kinematic features observed in the normalized distributions. Including only the one-loop running proportional to $\alpha_s$ as implemented in Ref.~\cite{Heinrich:2024rtg}, only the coefficient $C_{tH}$ of the non-zero coefficients in Tab.~\ref{benchmarks_invmass} is affected through its self-renormalization contribution. However, the resulting changes in the distributions of BM~2 and BM~5 are rather small, not changing their qualitative behaviour. They hence remain valid and distinctive kinematic benchmarks. Nonetheless, as shown in Ref.~\cite{DiNoi:2024ajj} for some scenarios with several operators and upon inclusion of further RGE effects, the effects can become phenomenologically relevant.

\subsection{Benchmarks under Truncation}
\begin{figure}[h!]
    \centering
    \begin{subfigure}{0.49\linewidth}
        \centering
        \includegraphics[width=\linewidth]{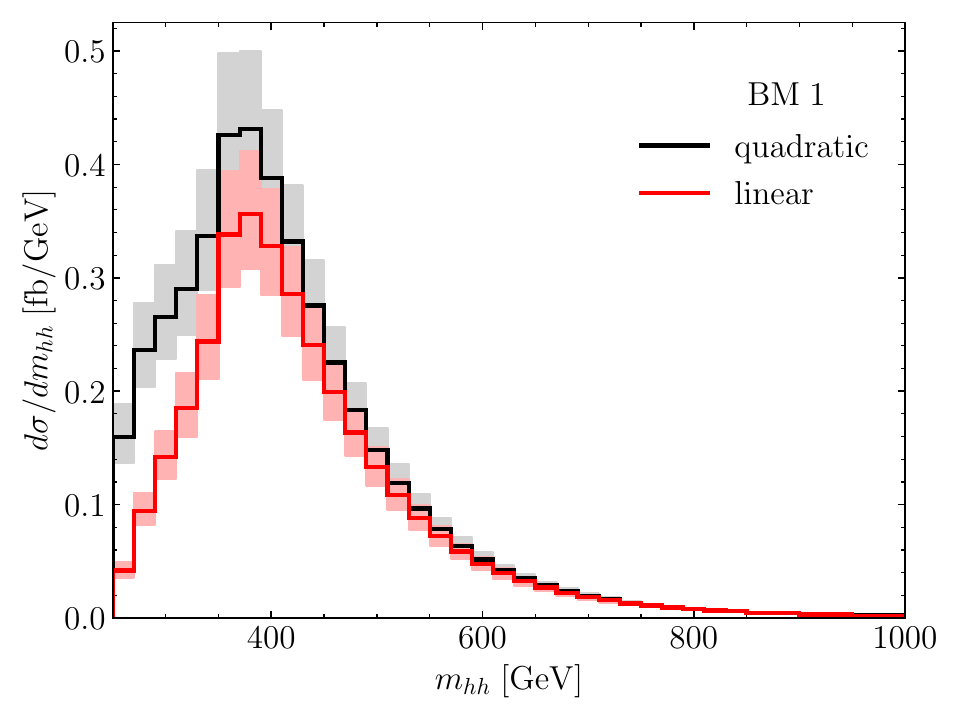}
    \end{subfigure}
    \hfill
    \begin{subfigure}{0.49\linewidth}
        \centering
        \includegraphics[width=\linewidth]{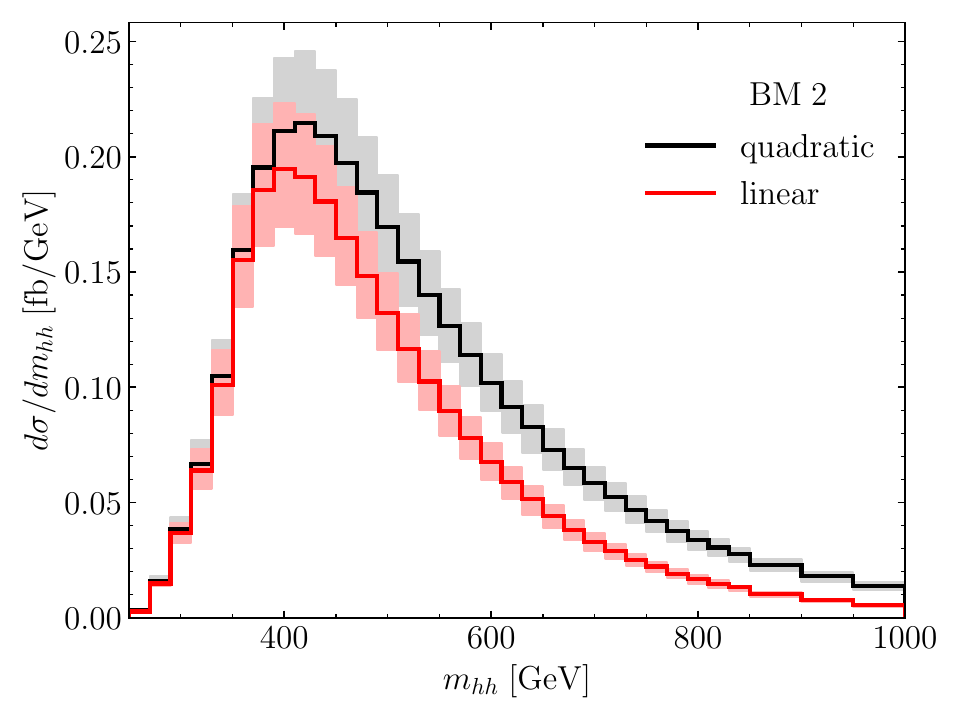}
    \end{subfigure}    
    \vspace{0.5cm}
    \centering
    \begin{subfigure}{0.49\linewidth}
        \centering
        \includegraphics[width=\linewidth]{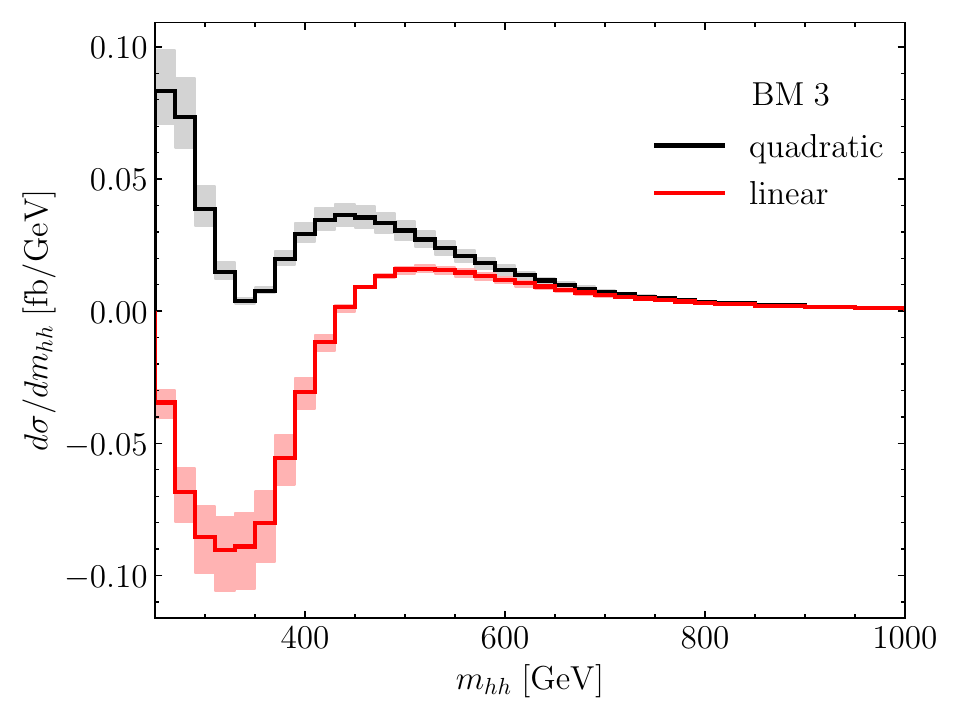}
    \end{subfigure}
    \hfill
    \begin{subfigure}{0.49\linewidth}
        \centering
        \includegraphics[width=\linewidth]{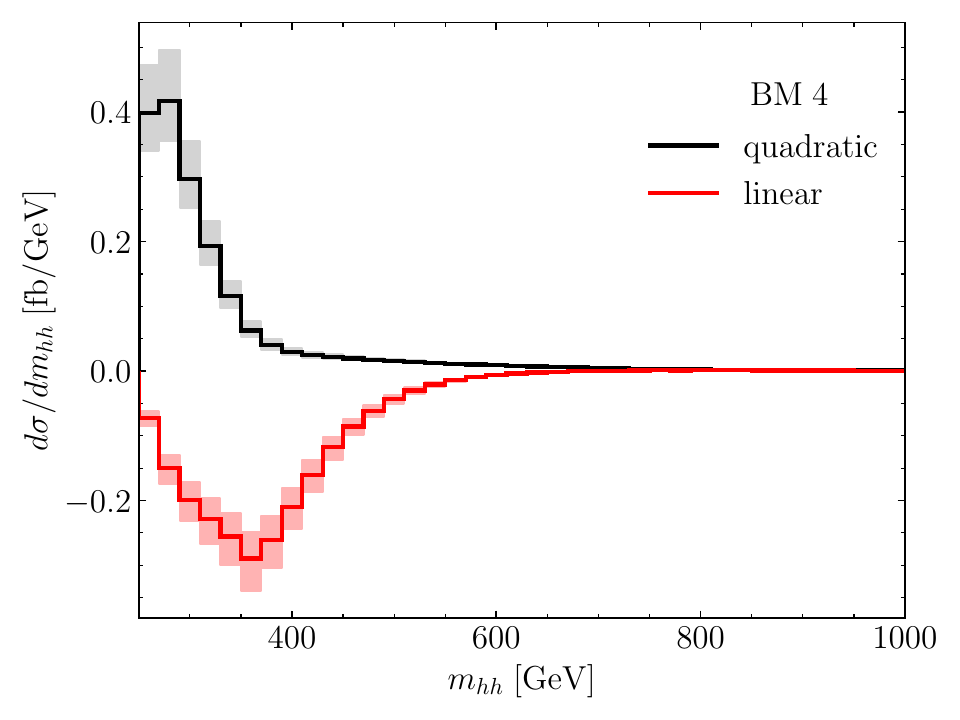}
    \end{subfigure}    
    \vspace{0.5cm}
    \begin{subfigure}{0.49\linewidth}
        \centering
        \includegraphics[width=\linewidth]{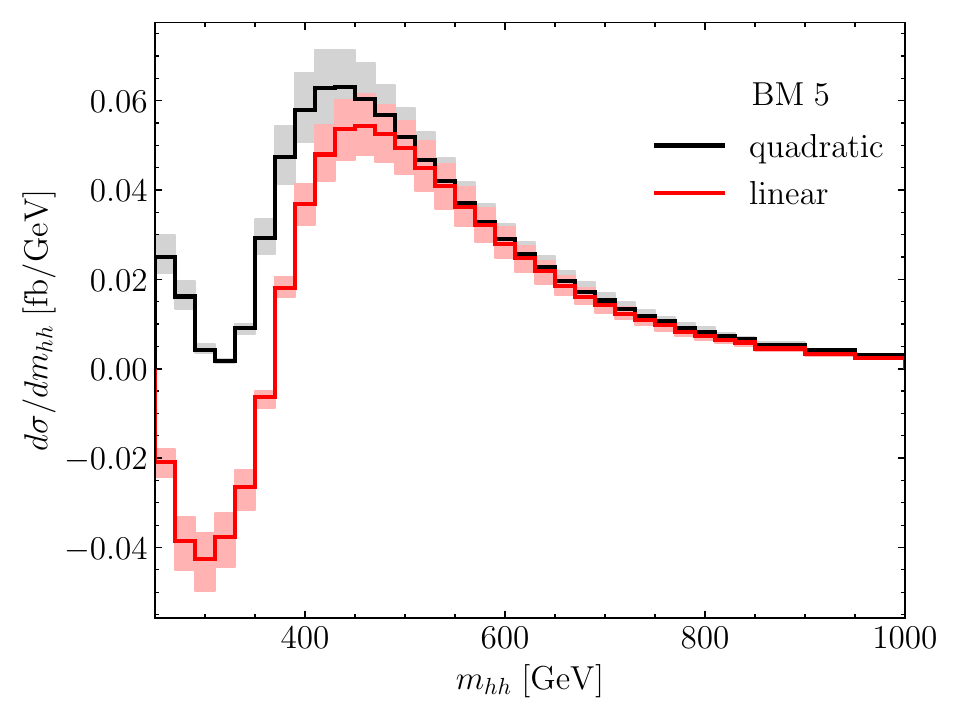}
    \end{subfigure}
    \caption{Comparison of linear and quadratic truncation in the invariant mass distribution for benchmark points 1-5. In light gray (red) we show the envelope of a three-point scale variation of the quadratic (linear) distributions.}
    \label{benchmarks_truncation}
\end{figure}

In the following, we examine how the benchmark points obtained with quadratic contributions behave under changes of the truncation procedure. Fig.~\ref{benchmarks_truncation} displays the invariant mass distributions under quadratic and linear truncation, including the envelope of the three-point scale variation $\mu \in \{\overline{\mu}/2, \overline{\mu}, 2 \overline{\mu} \}$ around the central scale $\overline{\mu} = m_{hh}/2$. 
While only benchmark points 1 and 2 yield non-negative invariant mass distributions also under linear truncation and their kinematic features remain essentially unchanged, benchmark points 3-5 lead to drastic shape changes and negative bins for small invariant masses. We verified that this is not only true for our selection of the benchmark points 3-5, but really holds for any set of Wilson coefficients that reproduces similar kinematic features in quadratic truncation.\footnote{The validation has been performed using the clustering algorithm described in Ref.~\cite{Brivio:2025sib}.} Furthermore, the kinematic features of benchmark points 3-5 cannot be reproduced within the linear truncation scheme, such that only benchmark points 1 and 2 are suitable for an analysis employing linear truncation. In addition, this might suggest that for these regions of the SMEFT parameter space the linear truncation scheme is an inadequate approximation, or that the SMEFT expansion is not well-behaved. We notice that the negative cross section bins are not just an artifact of a small SM prediction (and hence small interference) as they extend also to the bins where the SM cross section is maximal. The necessity of the inclusion of the quadratic terms for a positive cross section is a hint that also the inclusion of dimension-8 operators for concrete UV scenarios might be necessary. 
Consequently, an experimental indication for the realization of the benchmark points 3–5 would favor an interpretation in terms of HEFT, or require the inclusion of dimension-8 contributions \cite{Brivio:2026}. As shown in Ref.~\cite{Grober:2025vse}, under simplified assumptions for Higgs-pair production, possible differences between HEFT and SMEFT are ultimately a question of the convergence of the EFT expansion. It would be interesting to investigate whether benchmark scenarios 3--5 can be realized in a concrete UV theory. 

\subsection{Transverse Momentum Distributions}
\begin{figure}[ht]
    \centering
    \includegraphics[width=1.0\linewidth]{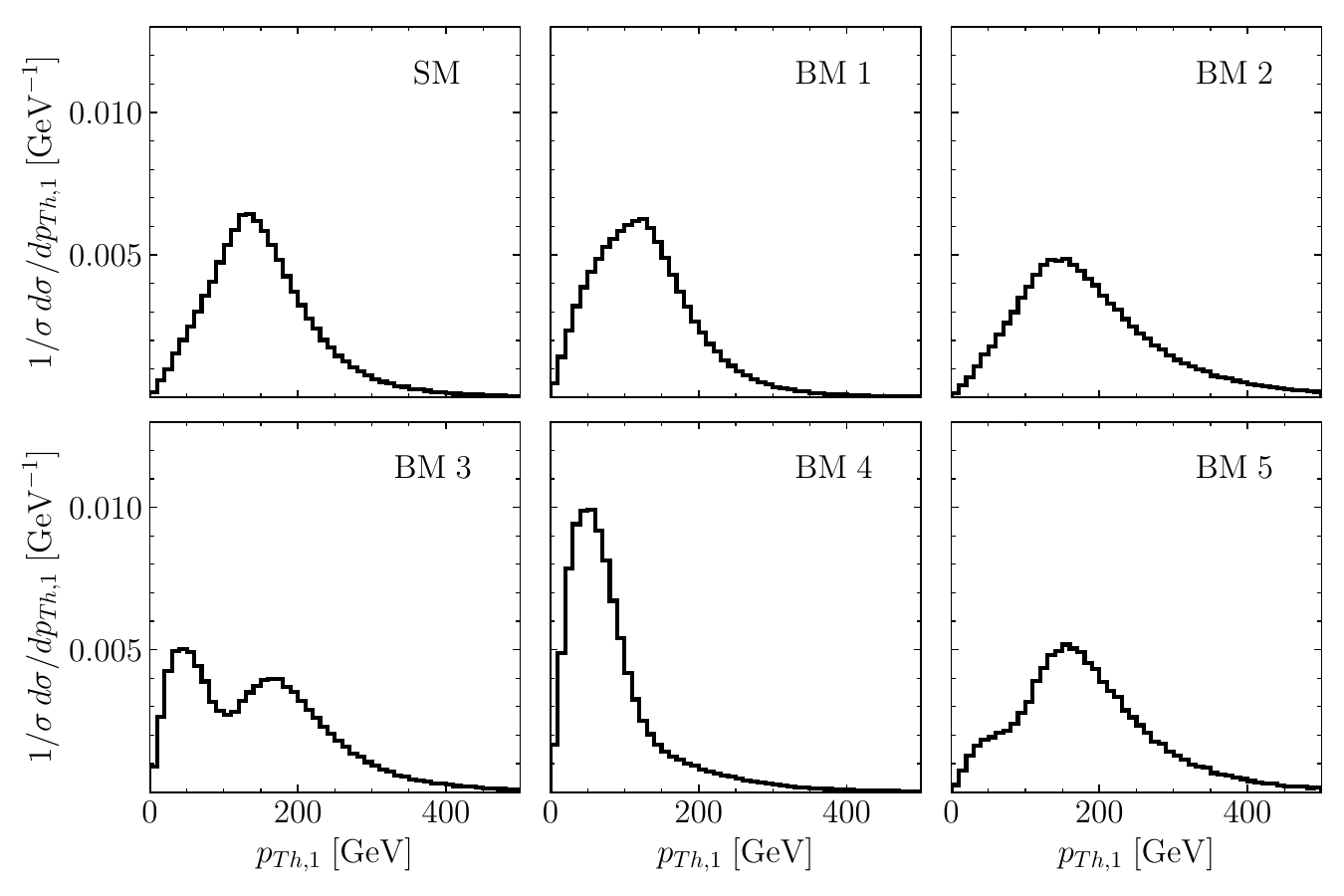}
    \caption{Normalized leading transverse momentum distributions of benchmark points 1-5 from Tab.~\ref{benchmarks_invmass}. The shapes for the subleading transverse momentum only deviate marginally and are therefore not shown explicitly.}
    \label{benchmarks_pTH1}
\end{figure}

\begin{figure}[h!]
    \centering
    \begin{subfigure}{0.49\linewidth}
        \centering
        \includegraphics[width=\linewidth]{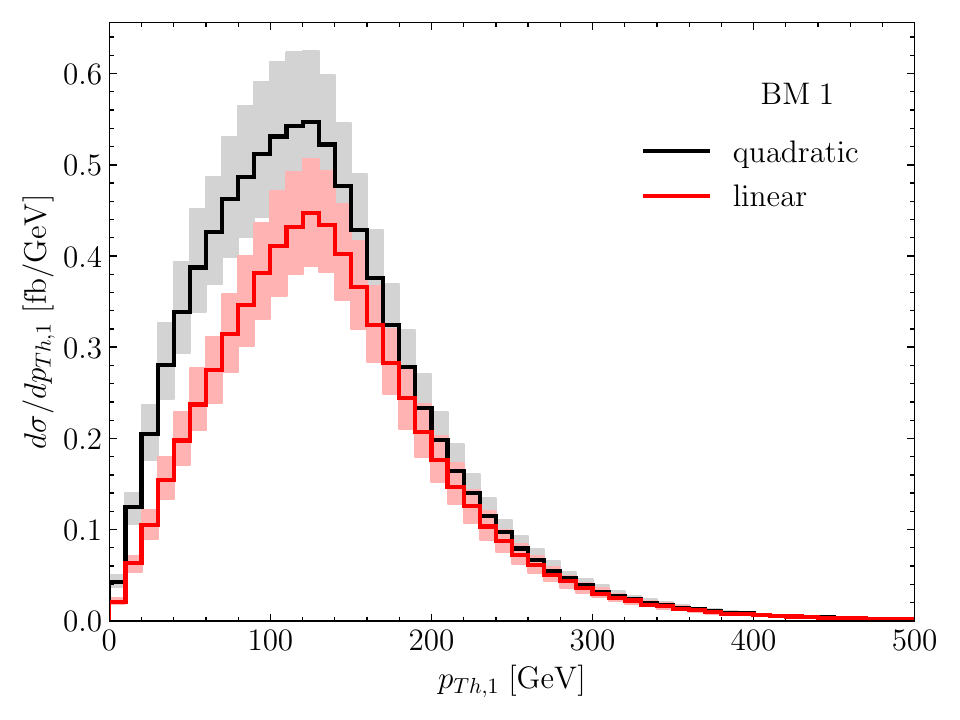}
    \end{subfigure}
    \hfill
    \begin{subfigure}{0.49\linewidth}
        \centering
        \includegraphics[width=\linewidth]{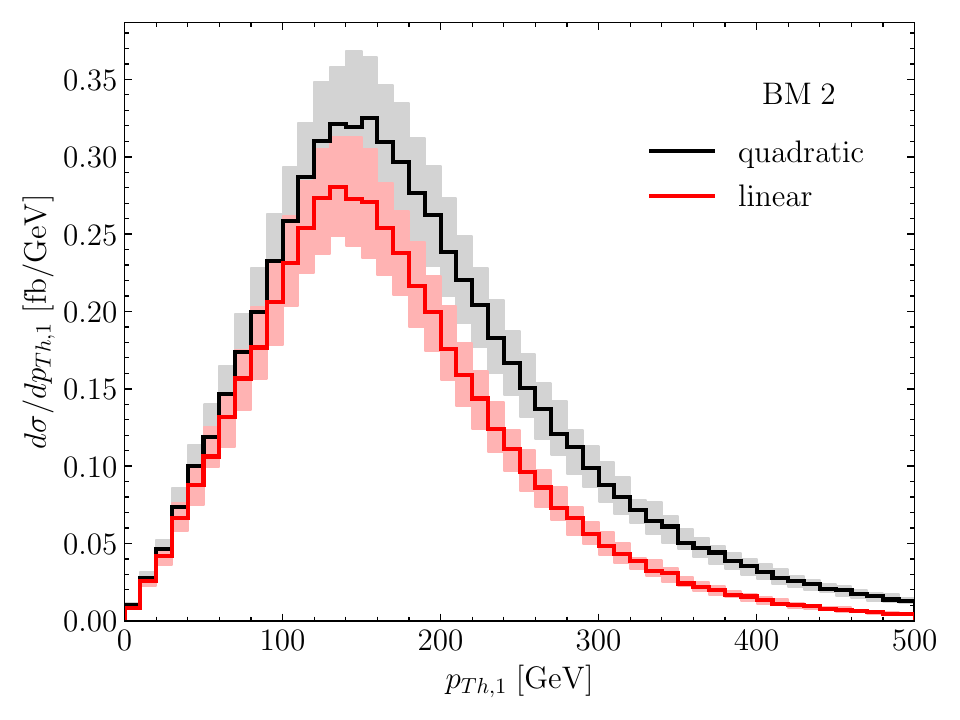}
    \end{subfigure}    
    \vspace{0.5cm}
    \centering
    \begin{subfigure}{0.49\linewidth}
        \centering
        \includegraphics[width=\linewidth]{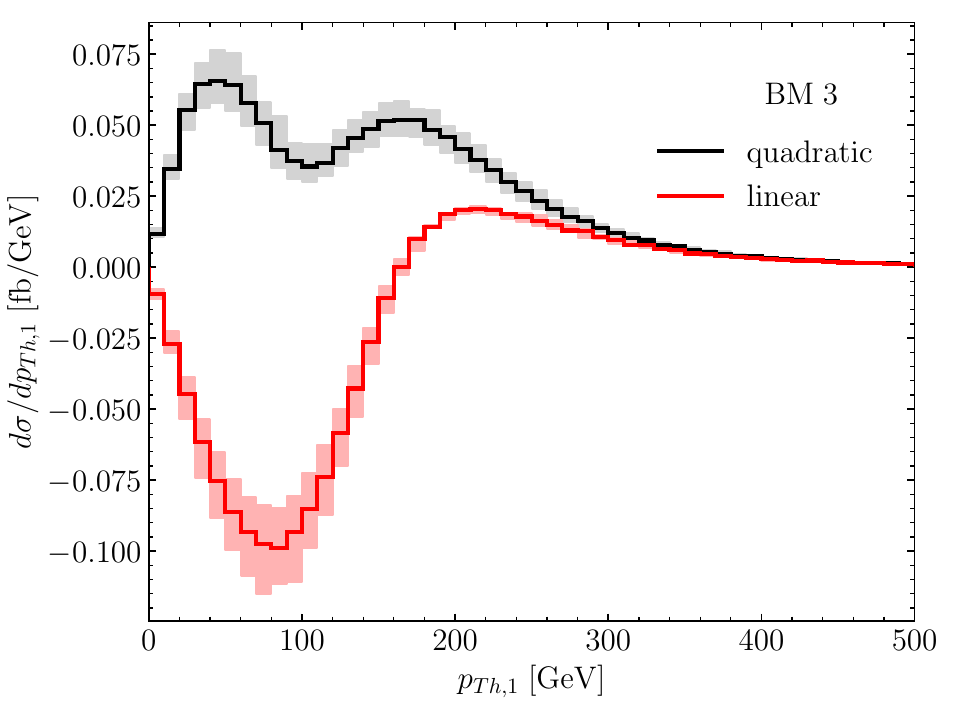}
    \end{subfigure}
    \hfill
    \begin{subfigure}{0.49\linewidth}
        \centering
        \includegraphics[width=\linewidth]{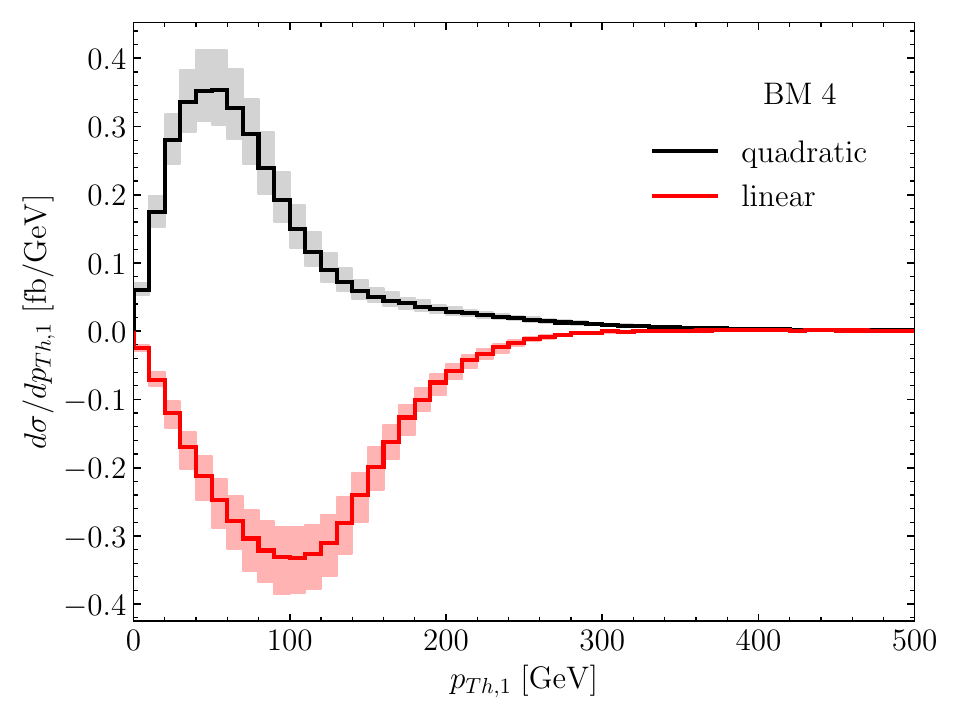}
    \end{subfigure}    
    \vspace{0.5cm}
    \begin{subfigure}{0.49\linewidth}
        \centering
        \includegraphics[width=\linewidth]{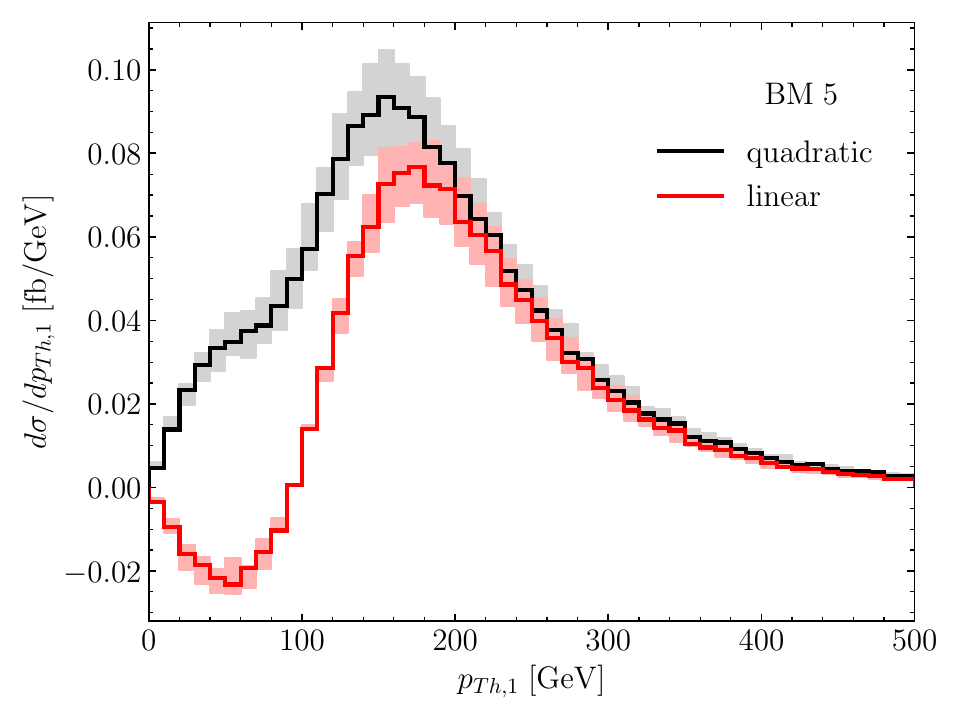}
    \end{subfigure}
    \caption{Comparison of linear and quadratic truncation in the leading transverse momentum distributions of benchmark points 1-5. In light gray (red) we show the envelope of a three-point scale variation of the quadratic (linear) distributions.}
    \label{benchmarks_truncation_pTH1}
\end{figure}

For our analysis of the transverse momentum spectra, we show the corresponding distributions of the leading transverse momentum for the benchmark points listed in Tab.~\ref{benchmarks_invmass} in Fig.~\ref{benchmarks_pTH1}. We do not display the subleading transverse momentum distributions, as they exhibit only marginal differences with respect to the leading transverse momentum spectra. Similarly to the $m_{hh}$ shapes, the transverse momentum shapes also show significant shape modifications for the benchmark points 3–5: we observe a double-peak structure for BM 3, a pronounced peak that is shifted towards $p_{Th, i} \simeq 50 \, \mathrm{GeV}$ for BM 4, and a left shoulder for BM 5. Again, these kinematic features are primarily driven by the value of $C_H$. To further investigate this behavior under truncation, we study the same distributions within the linear truncation approach in Fig.~\ref{benchmarks_truncation_pTH1}. As already observed for the invariant mass distributions, benchmark points 3–5 again lead to negative bins in the transverse momentum spectra. 

\newpage
\section{Conclusions} \label{sec_conclusions}
In this article, we studied di-Higgs production including the effect of dimension-6 SMEFT operators, together with NLO QCD corrections. We parameterize the results for the inclusive di-Higgs cross section and the invariant mass, leading and subleading transverse momentum distributions in terms of a set of $A_i$-coefficients, whose numerical values we provide in the supplementary material, including three-point scale variations. \\
Looking toward the integrated luminosity of $1\,\mathrm{ab}^{-1}$ targeted in Report 5, the increased statistical reach of di-Higgs measurements will make differential information accessible, which will be key for SMEFT interpretations. Different combinations of Wilson coefficients produce similar inclusive cross sections, while leading to distinct distortions of the $m_{hh}$ and $p_{T,h}$ distributions. At the same time, only a limited number of shape classes may be experimentally distinguishable at the available precision. Our NLO parameterizations and shapes can be used by the experimental collaborations to set limits on signals with distinct kinematic features as already done for the HEFT~\cite{CMS:2025ngq} and ultimately improve constraints on SMEFT Wilson coefficients.
\\
In our phenomenological analysis, we considered two different truncation schemes for the SMEFT expansion: linear truncation that strictly neglects all terms of order $\mathcal{O}(\Lambda^{-4})$ and quadratic truncation, which additionally retains the terms quadratic in the dimension-6 Wilson coefficients, arising from the square of the dimension-6 amplitude. For the latter, we identified five benchmark scenarios for the normalized invariant mass distributions, which are mainly driven by the value of the $C_H$ coefficient, and exhibit kinematic features that deviate significantly from the SM prediction. At the same time, they show qualitative similarities to the results of previous HEFT studies. In contrast, when adopting the linear truncation approach, benchmark points 3-5 yield absolute invariant mass distributions with negative bins, suggesting 
the insufficiency of linear truncation or a breakdown of the EFT expansion within these regions of the parameter space. A similar behavior is observed for the normalized transverse momentum distributions: only benchmark points 3-5 induce sizable shape changes, which, under linear truncation, again lead to unphysical negative bins. \\
In conclusion, a SMEFT analysis of di-Higgs production should consider both truncation schemes as the associated truncation uncertainties can become substantial. It remains to be understood whether these effects merely indicate that the linear truncation provides an inadequate approximation in certain regions of the SMEFT parameter space, or whether they are instead unphysical as the EFT expansion breaks down. In particular, addressing this question requires a complete understanding of the dimension-8 contributions, as well as a comprehensive comparison with realistic UV realizations, which have been matched to the SMEFT up to dimension-8.

\section*{Acknowledgements}
We would like to thank Alejo Rossia for useful discussions and Ilaria Brivio for collaboration in the early stages of this project. \\
The research of BC and GH is supported by the Deutsche Forschungsgemeinschaft (DFG, German Research Foundation) under grant 396021762 - TRR 257. The work of RG is supported by a STARS@UniPD grant under the acronym ``HiggsPairs'' and in part by the Italian MUR Departments of Excellence grant 2023-2027 ”Quantum Frontiers”. The authors acknowledge support from the COMETA COST Action CA22130.

\appendix
\section{Parameterization of the Results} \label{appendix_parameterization}
In this appendix, we present the full parametrization of the inclusive di-Higgs cross section in linear and quadratic truncation as defined in Eqs.~(\ref{param_lin_incl}) and (\ref{param_quad_incl}) to fix the conventions also employed in the ancillary file. Note that, while we write the following relations in terms of the inclusive quantities, the numbering is equivalent for the differential $A_i$-coefficients entering in Eqs.~(\ref{param_lin_diff}) and (\ref{param_quad_diff}). We write the total cross section under linear truncation as
\begin{align}
    \sigma_{\rm SMEFT}\big|_{\rm lin} = A_0^{\rm incl} + A_1^{\rm incl} C_{H, \rm kin} + A_2^{\rm incl} C_H + A_3^{\rm incl} C_{tH} + A_4^{\rm incl} C_{HG} + A_5^{\rm incl} C_{tG} + A_6^{\rm incl} C_{Qt}^{(1)} + A_7^{\rm incl} C_{Qt}^{(8)} \, ,
\end{align}
such that $A_0^{\rm incl}$ is the SM cross section including NLO QCD contributions. Instead, the quadratic truncation cross section contains additional terms proportional to the product of two dimension-6 Wilson coefficients, i.e.
\begin{align}
    \sigma_{\rm SMEFT}\big|_{\rm quad} =  \sigma_{\rm SMEFT}\big|_{\rm lin} + \sum_{\substack{i,j \\ i \le j}} A_{ij}^{\rm incl} \, C_i^{(6)} C_j^{(6)} \, ,
\end{align}
where the latter sum is expanded as

%\begin{equation}
%\begin{aligned}
%    \sum_{\substack{i,j \\ i \le j}} A_{ij}^{\rm incl} \, C_i^{(6)} C_j^{(6)} &= A_8^{\rm incl} C_{H, \rm kin}^2 + A_9^{\rm incl} C_H^2 + A_{10}^{\rm incl} C_{tH}^2 + A_{11}^{\rm incl} C_{HG}^2 + A_{12}^{\rm incl} C_{tG}^2 + A_{13}^{\rm incl} (C_{Qt}^{(1)})^2 + A_{14}^{\rm incl} (C_{Qt}^{(8)})^2 \\
%    &+ A_{15}^{\rm incl} C_{H, \rm kin} C_H + A_{16}^{\rm incl} C_{H, \rm kin} C_{tH} + A_{17}^{\rm incl} C_{H, \rm kin} C_{HG} + A_{18}^{\rm incl} C_{H, \rm kin} C_{tG} \\
%    &+ A_{19}^{\rm incl} C_{H, \rm kin} C_{Qt}^{(1)} + A_{20}^{\rm incl} C_{H, \rm kin} C_{Qt}^{(8)} + A_{21}^{\rm incl} C_H C_{tH} + A_{22}^{\rm incl} C_H C_{HG} \\
%    &+ A_{23}^{\rm incl} C_H C_{tG} + A_{24}^{\rm incl} C_H C_{Qt}^{(1)} + A_{25}^{\rm incl} C_H C_{Qt}^{(8)} + A_{26}^{\rm incl} C_{tH} C_{HG} + A_{27}^{\rm incl} C_{tH} C_{tG} \\
%    &+ A_{28}^{\rm incl} C_{tH} C_{Qt}^{(1)} + A_{29}^{\rm incl} C_{tH} C_{Qt}^{(8)} + A_{30}^{\rm incl} C_{HG} C_{tG} + A_{31}^{\rm incl} C_{HG} C_{Qt}^{(1)} + A_{32}^{\rm incl} C_{HG} C_{Qt}^{(8)} \\
%    &+ A_{33}^{\rm incl} C_{tG} C_{Qt}^{(1)} + A_{34}^{\rm incl} C_{tG} C_{Qt}^{(8)} + A_{35}^{\rm incl} C_{Qt}^{(1)} C_{Qt}^{(8)} \, .
%\end{aligned}
%\end{equation}

\begin{equation}
\begin{aligned}
    \smash{\sum_{i,j \geq i}} A_{ij}^{\rm incl} \, C_i^{(6)} C_j^{(6)} \; &= \; A_8^{\rm incl} C_{H, \rm kin}^2 \; + \; A_9^{\rm incl} C_H^2 \; + \; A_{10}^{\rm incl} C_{tH}^2 \; + \; A_{11}^{\rm incl} C_{HG}^2 \; + \; A_{12}^{\rm incl} C_{tG}^2 + \; A_{13}^{\rm incl} (C_{Qt}^{(1)})^2 \\
    &+ \; A_{14}^{\rm incl} (C_{Qt}^{(8)})^2 \; + \; A_{15}^{\rm incl} C_{H, \rm kin} C_H \; + \; A_{16}^{\rm incl} C_{H, \rm kin} C_{tH} \; + \; A_{17}^{\rm incl} C_{H, \rm kin} C_{HG} \\
    &+ \; A_{18}^{\rm incl} C_{H, \rm kin} C_{tG} \; + \; A_{19}^{\rm incl} C_{H, \rm kin} C_{Qt}^{(1)} \; + \; A_{20}^{\rm incl} C_{H, \rm kin} C_{Qt}^{(8)} \; + \; A_{21}^{\rm incl} C_H C_{tH} \\
    &+ \; A_{22}^{\rm incl} C_H C_{HG} \; + \; A_{23}^{\rm incl} C_H C_{tG} \; + \; A_{24}^{\rm incl} C_H C_{Qt}^{(1)} \; + \; A_{25}^{\rm incl} C_H C_{Qt}^{(8)} \; + \; A_{26}^{\rm incl} C_{tH} C_{HG} \\
    &+\; A_{27}^{\rm incl} C_{tH} C_{tG} \; + \; A_{28}^{\rm incl} C_{tH} C_{Qt}^{(1)} \; + \; A_{29}^{\rm incl} C_{tH} C_{Qt}^{(8)} \; + \; A_{30}^{\rm incl} C_{HG} C_{tG} \; + \; A_{31}^{\rm incl} C_{HG} C_{Qt}^{(1)} \; \\
    &+\; A_{32}^{\rm incl} C_{HG} C_{Qt}^{(8)} \;+\; A_{33}^{\rm incl} C_{tG} C_{Qt}^{(1)} \; + \; A_{34}^{\rm incl} C_{tG} C_{Qt}^{(8)} \; + \; A_{35}^{\rm incl} C_{Qt}^{(1)} C_{Qt}^{(8)} \, .
\end{aligned}
\end{equation}

\bibliographystyle{SciPost_bibstyle}
\bibliography{HHEFT}

\end{document}